\documentclass{elsart4}

\usepackage{graphicx}
\usepackage{dcolumn}
\usepackage{amsmath}
\usepackage{amssymb}
\begin{document}

\begin{frontmatter}
\vspace{-3cm}
\title{Magnetism in Carbon Structures\thanksref{tit1}}
\thanks[tit1]{This work has been done with the collaboration of  K.-H. Han, A.~Setzer,
D. Spemann, and T. Butz. Discussions with Y. Kopelevich and T. Makarova
are gratefully acknowledge. This work was supported by the DFG under
contract DFG ES 86/6. One  author (P.E.) acknowledges useful discussions
with K. Kusakabe and A.~Krasheninnikov, and M. Ziese for a carefully
reading of the manuscript.}

\author[aff1]{Pablo Esquinazi\corauthref{cor1}}
\ead{esquin@physik.uni-leipzig.de}
\author[aff1]{Roland~H\"ohne}
\corauth[cor1]{Tel:+49-341-9732751; fax:+49-341-9732769}
\address[aff1]{Division of Superconductivity and Magnetism,
University of Leipzig, Linn\'estrasse 5, D-04103 Leipzig, Germany}




\begin{abstract}
We discuss different magnetic phenomena observed in carbon-based
structures, in particular the diamagnetism, paramagnetism and
ferromagnetism observed in graphite, disordered carbon, fullerenes and
irradiated carbon structures.

\end{abstract}

\begin{keyword}
 Magnetic carbon \sep Soft magnetic materials \sep Ferromagnetic
semiconductors \sep Radiation effects \PACS 75.50.Pp \sep 75.50.Xx \sep
78.70.-g
\end{keyword}
\end{frontmatter}


\section{Introduction}\label{}
Within the fundamentals of magnetism we read in textbooks that the
magnetic order in solids below a critical temperature, like ferro-, ferri-
or antiferromagnetism, is characterized by a microscopic arrangement of
atomic magnetic moments. In order to account for ferromagnetism one needs,
for example, an exchange interaction of electrostatic origin between
localized magnetic moments of neighboring atoms or, in case of itinerant
electron magnetism, an effective exchange energy for the unpaired, no
longer localized electrons. Within these basic, well accepted concepts,
magnetic order was ``recognized" only for materials with 3$d$ or 4$f$
electrons. However, since 1991 we know that some organic molecules with
unpaired $\pi$-electrons show magnetic order at temperatures $T < 20~$K,
without having any metallic ions \cite{pimag}. We note also that
antiferromagnetic order in a $s,p$-electron system has been found below a
N\'eel temperature of $\sim 50~$K \cite{srdanov98}. We are not aware of
any law of physics that prohibits the existence of magnetic order in
systems with only $s,p$-electrons and at temperatures above room
temperature.

Not long time ago, carbon-based structures were basically accepted to be
diamagnetic or, in special cases, paramagnetic. Any ferromagnetic-like
signal was assumed to be due to magnetic impurities. No doubt, impurities
are a problem when the ``intrinsic" magnetic signals are relatively small.
However, it appears that instead of trying to study systematically  the
contribution of these impurities to the magnetic signals, the strategy of
``minimum work" based on prejudices was chosen, indirectly neglecting the
existence of interesting magnetic phenomena in carbon structures. In this
article we will review some recently published results that indicate that
we are at the beginning of our understanding of the magnetism in
carbon-based materials.

\section{Diamagnetism in oriented graphite}\label{dia}

Usually, a diamagnetic signal is measured in relatively ordered
carbon-based structures. The best example is oriented graphite, which
shows a large diamagnetic susceptibility. Measurements of highly oriented
pyrolytic graphite samples with rocking curve width at half maximum FWHM
$\le 0.4^\circ$ provide a value for the susceptibility $\chi_{||} = -(2.4
\pm 0.1) \times 10^{-5}$emu/gOe at 300~K and for fields applied parallel
to the $c-$axis (perpendicular to the graphene planes). The large
diamagnetism stems from ``fast-moving" electrons \cite{mcclure56} with a
small effective mass $m^* \sim 0.05 m_0$ (here $m_0$ is the free electron
mass). Within a tight-binding picture the small effective mass is due to
the large $\pi$-bonding overlap of the neighboring C-atoms in a single
layer given by the band parameter $\gamma_0 \propto 1/m^*$. For the
calculation of $\chi$ the linear dispersion relation for two-dimensional
(2D) graphene $E(k) \propto k$ was assumed \cite{wallace47}. On the other
hand, properties of massless ``Dirac" fermions are rediscovered nowadays
in the literature and are thought to play also a role in the observed
metal-insulator-like transition in the electrical resistivity
\cite{yakovadv03}. The main contribution to the $T-$dependence of
$\chi(T)$ at $T \gtrsim 150~$K  is given by $\partial f/\partial
E|_{E=E_F}$ ($f$ is the Fermi-Dirac distribution function) due to the
condensation of fermions into the Landau level with $n = 0$ after the
application of a magnetic field \cite{mcclure56}. Although experimental
results are in general in agreement with this theory, to explain the
saturation of $\chi(T)$ at low $T$ a series of additional band parameters
has to be included (see \cite{dresfibers} and Refs. therein). We note,
however, that this further developed theory does not account for the
measured anisotropy of the $g$-factors \cite{dresfibers}, a disagreement
that, to our knowledge, has not been solved yet. Moreover, high resolution
magnetization measurements in different HOPG samples indicate that $\chi$
has a shallow but well-defined and field dependent minimum at $T \lesssim
30~$K (as an example see Fig.~\ref{fig1}) in agreement with that reported
in \cite{yakov99}. This minimum is not due to magnetic impurities. The
origin for this minimum is not yet clarified and no attempt has been done
to check whether it is compatible with more elaborate theories for the
diamagnetism in graphite.

Due to the available sample quality, the measurement of the anisotropy in
the carrier diamagnetism of graphite is difficult. Taking into account
recent measurements of the electrical anisotropy that indicates a ratio
(parallel to perpendicular to the $c-$axis) of the order of or larger than
$10^4$ for well ordered samples, we expect to have a susceptibility
$|\chi_{\perp}|$ smaller than the atomic susceptibility of carbon in
graphite, which is of the order of $-5 \times 10^{-7}~$emu/gOe according
to literature. We have also to take internal misalignments of the
crystallites in the sample into account. At room temperature and for a
sample with FWHM $ = 0.4^\circ$, due to the $\chi_{||}$ component a value
of the order of $- 1.7 \times 10^{-7}$emu/gOe $\ll \chi_{\perp}$ would be
measured for fields applied nominally parallel to the planes. The effect
of this internal misalignment has been verified by electrical conductivity
measurements \cite{kempa03}.

\section{Paramagnetism in disordered carbon}\label{para}

\begin{figure}
\begin{center}
\includegraphics[width=0.5\textwidth]{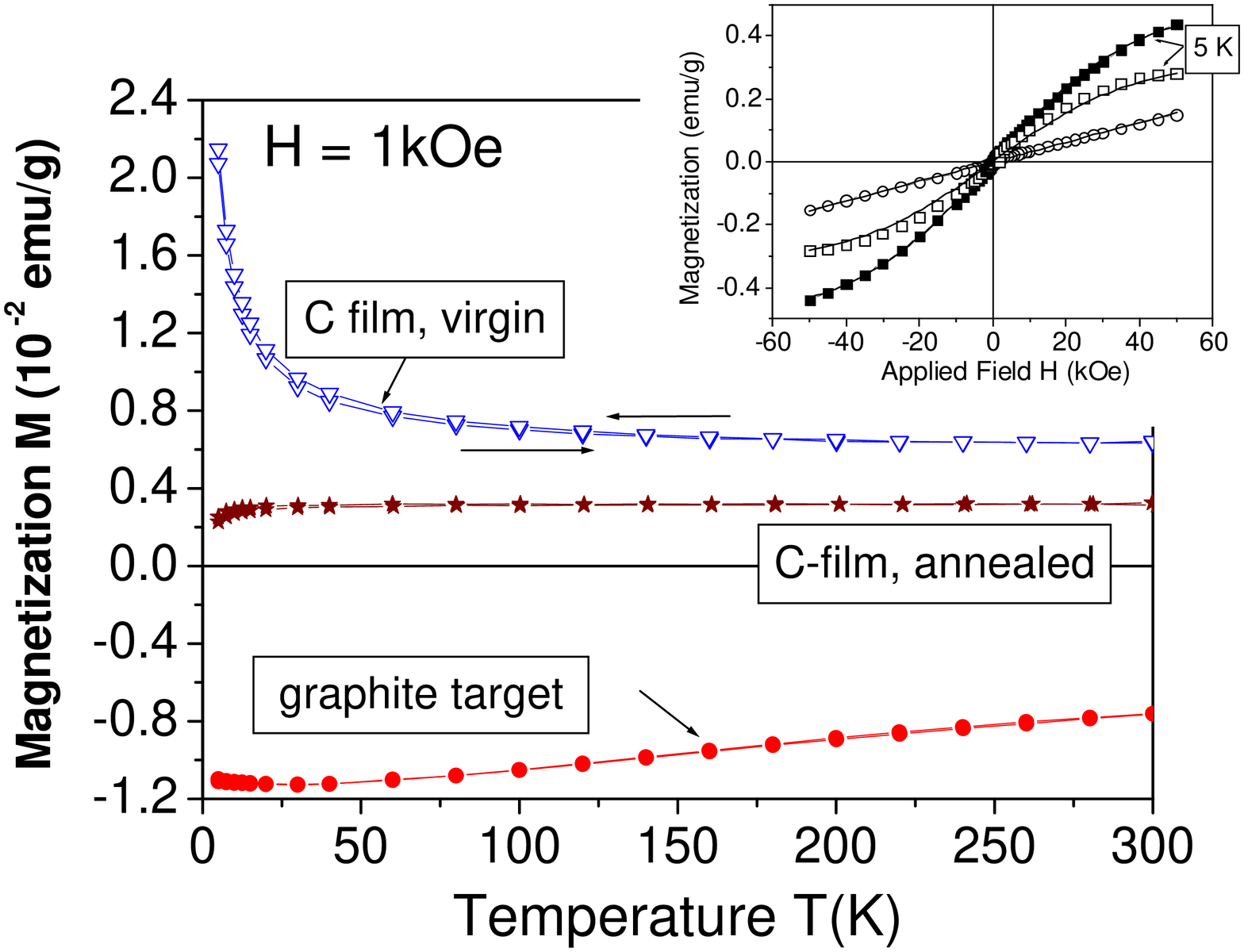}
\end{center}
\caption{Temperature dependence of the magnetization at 1~kOe obtained for
the graphite target made from an ultra pure graphite powder
\protect\cite{ringsdorf} and the films, before and after annealing at
1000~$^\circ$C for 10~h in vacuum. Inset: Field dependence of the
magnetization of the carbon films at 5 K ($\blacksquare$), and after
subtraction of the Pauli-like term ($\square$) given by the straight line
through the points (o) $M=3.1 \times 10^{-6} H$ emu/gOe. In the scale of
the figure the Pauli-like magnetization curves measured at 300 K (before
and after annealing) and at 5 K (after annealing) (o) practically
coincide. The continuous lines through the points ($\blacksquare,\square$)
are fits to the Langevin expression. Adapted from
\protect\cite{hohcfilms}.}\label{fig1}
\end{figure}
If we increase the disorder in a graphite structure the main diamagnetic
signal decreases in absolute value and turns to paramagnetic when the
disorder is large enough . A clear example can be seen in Fig.~\ref{fig1}
where the temperature dependence of the magnetization of a target made of
an ultra-pure graphite powder \cite{ringsdorf} and of the material
obtained from disordered carbon films prepared by pulsed laser deposition
(PLD) on Si substrates \cite{hohcfilms} are shown. Due to the random
alignment of the graphite crystallites in the graphite target, its
diamagnetic susceptibility at 300 K is a factor three smaller than
$\chi_{||}$ for HOPG. The data for the disordered carbon material indicate
a Curie-like dependence added to a Pauli-like contribution. The inset in
Fig.~\ref{fig1} shows the field dependence of the magnetization for the
carbon film material at 5~K. These data can be described by the usual
Langevin expression from which we obtain a spin density $N_s \sim
10^{19}$~g$^{-1}$. According to earlier work \cite{delhaes} the localized
spins in disordered carbon originate from unpaired electrons associated to
the existence of broken $\sigma$-dangling bonds. Interestingly, the
density of paramagnetic centers was correlated with the hydrogen
concentration in earlier work (see \cite{delhaes} and Refs. therein).
Annealing the disordered carbon material at 1000~$^\circ$C for 10 h in
vacuum reduces clearly the temperature dependent paramagnetic part, see
Fig.~\ref{fig1}. The observed behavior with annealing is not in favour of
a magnetic impurity contribution.

Within experimental error the hysteresis loops for the disordered carbon
films above 5~K are reversible. This indicates that a disordered mixture
of sp$^2$-sp$^3$ bonds, which exists in the disordered carbon films, does
not trigger automatically ferromagnetism. The overall behavior as well as
the spin density obtained for the disordered films prepared by PLD are
very similar to those obtained for activated carbon fibers \cite{shiba},
results interpreted in terms of antiferromagnetically interacting spins
from $\pi$-electrons originated at the edges of nanographite layers. One
may doubt, however, on the existence of edge states of graphene layers in
disordered carbon films. We note, however, that amorphous-like carbon
prepared from targets with a large hydrogen/carbon ratio was reported to
be ferromagnetic with a magnetization at saturation as large as 2~emu/g at
room temperature \cite{murata91}.

\section{Ferromagnetism in carbon-based structures}\label{fer}
There is a relatively large list of reports on organic ferromagnets since
in 1991 two metal-ion free compounds with Curie temperatures of 0.65~K
\cite{kino} and 16~K \cite{alle91} were discovered. However, the research
on the possible magnetic order in carbon-based structures appears to be
older; for a recent review on magnetism in carbon-based materials in which
most of the old work is included as well as a list of recent references on
organic magnets, see \cite{makareview}. In this contribution we would like
to point out a few experimental facts that indicate the intrinsic,
impurity free origin of the ferromagnetism in carbon structures with a
Curie temperature above room temperature.

\subsection{Impurity measurements}

A systematic and full characterization of the magnetic impurity content in
each of the samples is of primary importance and absolutely necessary.
Part of the reason for the weak interest attracted by the
ferromagneticlike hysteresis loops found in HOPG samples
\cite{yakovjltp00} as well as by older publications on magnetic carbon is
probably due to unclear impurity characterization. To improve this
situation  radically we have used the method of ``Particle Induced X-ray
Emission" (PIXE) for the impurity characterization. Having a carbon
matrix, this method reaches a sensitivity of $\sim 0.1~$ppm for all
magnetic metallic impurities and therefore is very convenient for all
relevant impurity analysis, prior and after any further characterization
with MFM and SQUID and sample handling. The accelerator LIPSION in Leipzig
with a proton micro-beam of 2.25 MeV energy and a broad-beam ($\sim
0.8~$mm diameter) of 2.0 MeV energy are used for the

PIXE measurements. It allows simultaneously proton irradiation and element
analysis (using the same protons). In comparison with the broad-beam, the
micro-beam has the advantage that enables us to make a detailed element
map of the sample within a depth of $\sim 46~\mu$m and check for
inhomogeneous distributions of impurities. Typical spectra and element
maps obtained in HOPG and fullerene samples as well as further details on
the used nuclear nanoprobe method can be found in
\cite{spemann03,pablopt,butz}.

\begin{figure}
\begin{center}
\includegraphics[width=0.5\textwidth]{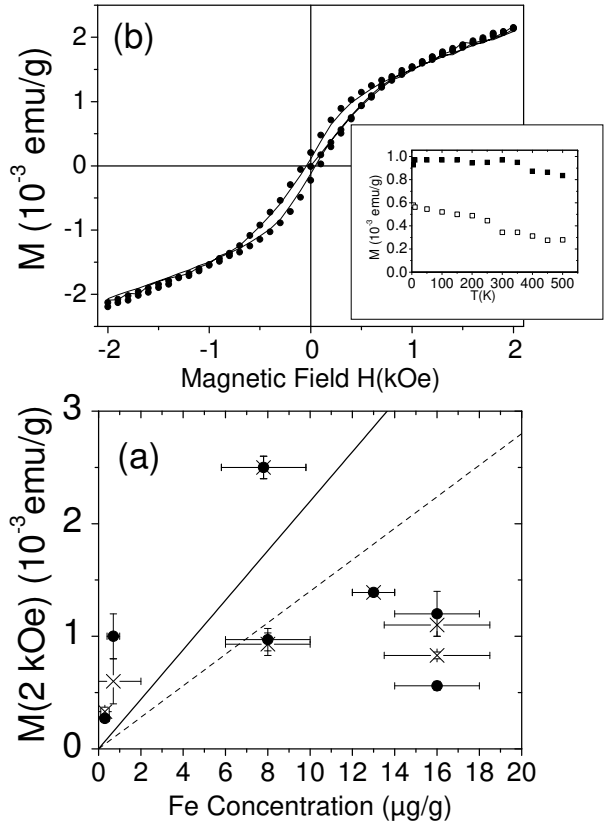}
\end{center}
\vspace{0.0cm} \caption{(a)Magnetization at 2 kOe as a function of Fe
concentration at two temperatures (($X$): 300~K, ($\bullet$): 5~K) for
different HOPG and Kish graphite samples. The solid line represents the
expected magnetization if Fe contained in the samples was in a
ferromagnetic state. The dashed line represents the analogous relation for
Fe$_3$O$_4$. (b) Hysteresis loops at 5~K ($\bullet$) and 300~K (continuous
line) of a HOPG sample (FWHM $\sim 1.3^\circ$) for fields applied parallel
to the graphene planes. No background was subtracted from the data. The
inset shows the $T-$dependence of the magnetization at saturation
$(\blacksquare)$ and its remanence $(\square$) after annealing 16~h at
700~K. Adapted from \protect\cite{pabloprb02}.} \label{fig2}
\end{figure}

\subsection{Ferromagnetic and paramagnetic signals
in highly oriented pyrolytic graphite}

For applied fields parallel to the graphene planes one is able to measure
ferro- and/or paramagnetic signals in HOPG samples. Usually those signals
would be overwhelmed by the diamagnetic one in the other field direction.
In Ref.~\cite{pabloprb02} we have studied in detail several HOPG samples
from different sources with concentration of Fe-impurities between $<
0.3~\mu$g/g to $\sim 19~\mu$g/g. We note that in those samples Fe is the
main magnetic impurity (other magnetic metallic impurities are below
2~$\mu$g/g). Figure~\ref{fig2}(a) shows that the magnetization at 2~kOe,
after substraction of background contributions, does not show a
correlation with the Fe-concentration. The results also indicate that the
ferromagnetic like hysteresis loops are weakly temperature dependent
between 5~K and 300~K, see inset in (b). A naive estimate assuming that
the Fe concentration behaves as Fe- or Fe$_3$O$_4$-bulk in the carbon
matrix would give the two lines shown in Fig.~\ref{fig2}(a). Except for
three HOPG samples the other show magnetization values below the ones
expected from those Fe-lines. This figure clearly shows how small the
ferromagnetic signals are that we are working with, as well as how much
care is needed for the sample handling in all these studies. However, the
assumption that such a small amount of Fe distributed in the carbon matrix
behaves ferromagnetically is neither consistent with the behavior we
observed in graphite samples with much larger Fe concentrations
\cite{pabloprb02}, nor with the recently obtained $T$-dependent,
paramagnetic behavior in Fe irradiated graphite \cite{hohneFe} nor in,
e.g., 4$d-5d$ metals, where paramagnetism as well as spin glass behaviors
are measured for Fe concentrations of $\sim 100~$ppm or larger
\cite{thomas}.

Figure~\ref{fig2}(b) shows two hysteresis loops for a HOPG sample. We note
that in this field range no sign of diamagnetism is measured but a
paramagnetic behavior added to the hysteresis. This is one of some
fortuitous cases where the diamagnetic component is small enough due to
the small misalignment of the sample in the SQUID.  The observed behavior
in Fig.~\ref{fig2} is in clear contrast with the T-dependent paramagnetic
contribution measured in samples with a larger amount of Fe-impurities
\cite{pabloprb02,hohneFe}.The reason for the temperature independent
(between 5 and 300~K) Pauli-like paramagnetism is probably related to the
intrinsic lattice disorder as in disordered carbon samples (see
Fig.~\ref{fig1}). The results shown in Fig.~\ref{fig2} belong to a
relatively disordered HOPG sample with FWHM$ \simeq 1.3^\circ$ and a
Pauli-like susceptibility $\chi \sim 5 \times 10^{-7}~$emu/gOe,  smaller
than that obtained for disordered carbon $\chi \sim 3 \times
10^{-6}~$emu/gOe \cite{hohcfilms}. We expect that by increasing the
lattice disorder a Curie-like paramagnetic contribution will appear.

The overall results including those discussed in Sec.~\ref{proton} suggest
that special lattice disorder {\em and} the influence of a light atom like
hydrogen may be the origin for the observed ferromagnetism. Recent density
functional calculations indicate that H adsorption on vacancy dangling
bonds gives rise to a localized magnetic moment of the order of $2.3
\mu_B$ \cite{lehtinen04}. If these defects or just H-atoms at edge states
can spin polarize the flat bands of graphite \cite{maru04}, then not Fe
but a few tens of ppm H might be enough to trigger the magnetic order we
observe. We stress that the ferromagnetic-like signal as well as the
paramagnetism observed in  HOPG are sample dependent. We expect that
rather perfect samples annealed at high enough temperatures $(T \gtrsim
2500~^\circ$C) and cooled down in high vacuum should show weaker ferro-
and/or paramagnetic signals. Measurements on pure, specially annealed
graphite powder \cite{ringsdorf} support this expectation.

\subsection{Ferromagnetism in fullerenes}

A magnetically ordered phase below the Curie temperature $T_C = 16~$K was
discovered in the organic charge-transfer salt [TDAE]C$_{60}$ (TDAE =
tetrakis(dimethylamino)ethylene) \cite{alle91}. Afterwards, ferromagnetism
was induced in $C_{60}$ by photo-assisted oxidation, which remains up to
$T_C \simeq 800~$K \cite{murakami96}, results that were confirmed by
another group \cite{maka03}. Recently, ferromagnetic ordering in a
hydrofullerite C$_{60}$H$_{24}$  with $T_C > 300~$K has been found
\cite{antonov02}. Polymerization of C$_{60}$ at temperatures and pressures
near the cage collapse and graphitization of the anisotropic 2D
rhombohedral Rh-C$_{60}$ phase leads to ferromagnetism with $T_C \gtrsim
500~$K \cite{makanat01,wood,naroz}. In general, the total concentration of
magnetic impurities reported in those studies appears to be too low to
give rise to the observed magnetization. Nevertheless, we need further
evidence that supports a magnetic impurity independent ordering. We note
that in C$_{60}$H$_{24}$ the initial ferromagnetic state strongly weakens
after one year of storage, a result unlikely to come from magnetic
impurities \cite{antonov02}. Measurements done on different polymerized
samples with different Fe-concentration (up to 400$~\mu$g/g) show very
similar magnetization behavior (magnetization at saturation $M_s \simeq
150~$A/m $= 0.065~$emu/g). This fact as well as the influence of heat
treatment on the magnetic behavior \cite{hohne02} do not favour an
interpretation in terms of magnetic impurities. Furthermore, the small
variation of remanence and coercivity with temperature gives no indication
for magnetism of small particles, a behavior similar to that observed in
different magnetic carbon-based structures including HOPG.

Magnetic force microscopy measurements performed in two polymerized
fullerenes could resolve well defined magnetic domains \cite{hancar03}.
These magnetic domains were obtained in a region where the magnetic
impurity concentration was less than $1~\mu$g/g as characterized by PIXE
measurements \cite{hancar03,spemann03}. These studies also showed that
only $\sim 30\%$ of the pure sample area was magnetic, a result that
speaks for the inhomogeneous character of the ferromagnetism and against a
strong correlation between the magnetic ordering and rhombohedral
structure. Recently published band structure calculations of Rh-C$_{60}$
performed in the local-spin-density approximation found no magnetic
solution for Rh-C$_{60}$ and energy bands with different spins are found
to be identical and not split, concluding that the rhombohedral distortion
of C$_{60}$ itself cannot induce magnetic ordering in polymerized
fullerene  \cite{Boukhvalov}. These results are in agreement with the
inhomogeneous magnetic distribution measured by MFM and suggest that the
magnetic ordering is related to other carbon structures that are formed
before the fullerene cages collapse \cite{wood}. New theoretical and
experimental work \cite{barbara} suggest that hydrogen may play an
important role on the magnetic ordering found in fullerenes and is
something that one should carefully check in future studies, as the
experiments described below indicate.

\subsection{Magnetic ordering induced by proton
irradiation}\label{proton}

\begin{figure}
\begin{center}
\includegraphics[width=0.5\textwidth]{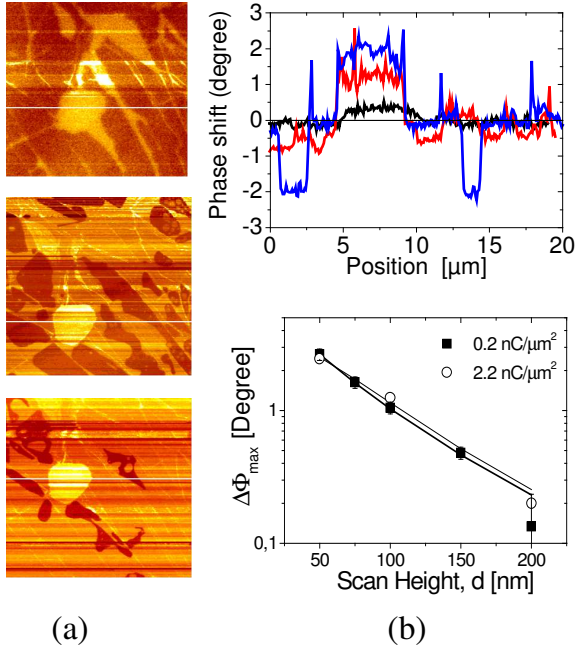}
\end{center}
\vspace{0.0cm} \caption{(a)Magnetic force gradient images $(20 \times 20
\mu$m$^2$) of a spot and its surroundings irradiated with
0.115~nC/$\mu$m$^2$ (proton current $I =171~$pA). The images were taken,
from top to bottom, before field application, after applying a field of
$\sim 1~$kOe in the $+z$ direction parallel to the $c$-axis, and in the
$-z$ direction. The tip-to-sample distance was 50nm. (b) Top: The
corresponding phase shift  obtained at the line scans (white straight
lines in figures (a)). The spot is located between $\sim 5 \mu$m and $\sim
10 \mu$m and the bottom, upper and middle lines in this region correspond
to measurements before and after application of a field in $-z$ and $+z$
direction. Adapted from Ref.~\protect\cite{han03}. Bottom: Scanning height
dependence of the maximum phase shift at proton irradiated spots with
fluences of 0.2~nC/$\mu$m$^2$ and 2.2~nC/$\mu$m$^2$. Solid lines are fits
with the point probe approximations. Adapted from
Ref.~\protect\cite{hanap}.} \label{fig3}
\end{figure}
Proton irradiation of graphite provides us with the unique possibility of
a complete analysis of the main magnetic elements in the sample and
simultaneously implant hydrogen and produce defects that may trigger
magnetic ordering in carbon structures. The measurements with the SQUID
and the MFM we have done show indeed clear signs of magnetic ordering
after irradiating HOPG samples \cite{pabloprl03,han03} and disordered
carbon films \cite{pablopt}. In the published literature on magnetism in
carbon-based structures one realizes that apparently hydrogen (or maybe
also other light atoms like oxygen) plays a role in the reported
ferromagnetism. Specially the work in Ref.~\cite{murata91} showed clearly
that the saturation magnetization of disordered carbon prepared from
different H-rich targets increases with the H-concentration of the
starting materials.

The results after proton irradiation leave no doubt that magnetic ordering
exists in a carbon structure without the influence of magnetic ions.
Neither the total amount of magnetic impurities is sufficient to account
for the measured magnetization nor the creation of magnetic spots in the
micrometer range with the proton micro-beam can be understood based on
magnetic metal-ion impurity concentration below 1~ppm as the PIXE results
indicate. An example of a magnetic spot and the field gradient response
after applying a magnetic field in different directions is shown in
Fig.~\ref{fig3}. We can clearly recognize the magnetic signal at the
bombardment position as well as magnetic ``structures" outside the spot,
which also change after application of an external magnetic field (note
that the MFM measurements were always done without applied field)
\cite{han03}. In the figure we show also how the magnetic signal depends
on the sample-tip height of the MFM. Using these data and different models
from literature we can estimate, although with a relatively large error, a
magnetization at the spot surface of the order of $\sim 10^6~$A/m $\sim
400~$emu/g \cite{hanap}.

\section{Open questions}
Further experimental characterization  (using the broad spectrum
of methods in magnetism research) and sample preparation studies
are necessary to understand and stabilize the magnetic ordering
found in carbon structures. The following questions are a short
list that should be clarified in the near future:
\begin{enumerate}
\item The role of H-atoms, implanted by irradiation as well as those
already in the sample. \item The contribution to the magnetic ordering
from lattice defects produced by irradiation and their possible influence
as H-trapping centers. \item The effective magnetic moment of magnetic
impurities in graphite as well as in disordered carbon structures. \item
The maximum achievable saturation magnetization in carbon structures.
According to theoretical work this might be as large as three times that
of pure Fe \cite{ovchi91}. \item The range of Curie temperature.
\end{enumerate}




\end{document}